\documentclass[pra,onecolumn,preprint]{revtex4}
\usepackage{graphicx}
\usepackage{amsfonts}

\begin{document}

\preprint{}
\title[Optical analogy to quantum Fourier transform]{Optical analogy to
quantum Fourier transform based on pseudorandom phase ensemble}
\author{Jian Fu, Wei Fang, Yongzheng Ye}
\affiliation{State Key Lab of Modern Optical Instrumentation, College of Optical Science
and Engineering, Zhejiang University, Hangzhou, 310027, China}

\begin{abstract}
In this paper, we introduce an optical analogy to quantum Fourier
tanformation based on a pseudorandom phase ensemble. The optical analogy
also brings about exponential speedup over classical Fourier tanformation.
Using the analogy, we demonstrate three classcial fields to realize Fourier
transform similar to three quantum particles.
\end{abstract}

\date{today}
\startpage{1}
\email{jianfu@zju.edu.cn}
\maketitle

\section{Introduction \label{sec1}}

Quantum Fourier transform is the most important tool of quantum computation 
\cite{Nielsen}, and one of the algorithms which can bring about exponential
speedup. Shor's algorithm, hidden subgroup problem and solving systems of
linear equations all make use of quantum Fourier transform \cite{Montanaro}.
Quantum Fourier transform utilizes the superposition of quantum state,
whereby the required time and space for computation can be notably reduced
from $2^{n}$ to $n$. Hence, the implementation of quantum Fourier algorithm
is crucial to exponential speedup in quantum computation \cite%
{Williams,Deutsch}.

A novel method to simulate quantum entanglement using classical fields
modulated with pseudorandom phase sequences was introduced in Ref. \cite{Fu1}%
, which can realize tensor product in quantum computation and the
representation of arbitrary quantum state \cite{Fu2}. At the same time, Ref. 
\cite{Fu3} also introduced a new concept, pseudorandom phase ensemble, to
simulate a quantum ensemble. However, it is necessary to implement similar
quantum Fourier transform algorithm to implement some current quantum
algorithms. Therefore, in this paper, we, at first utilize this method to
implement the simulation of quantum Fourier transform, then investigate the
required computational resources, and at last take three kinds of fields as
an example to verify this algorithm.

\section{Quantum Fourier transform \label{sec2}}

Generally, quantum Fourier transform takes as input a vector of complex
numbers, $f(0),f(1)\ldots ,f(N-1)$, and output a new vector of complex
numbers $\tilde{f}(0),\tilde{f}(1)\ldots ,\tilde{f}(N-1)$ as following:%
\begin{equation}
\tilde{f}\left( k\right) =\frac{1}{\sqrt{N}}\sum\limits_{j=0}^{N-1}e^{\frac{%
2\pi i}{N}jk}f\left( j\right) .  \label{eq1}
\end{equation}

This calculation involves the additions and multiplications of $N=2^{n}$
complex numbers, leading to an increase of computational complexity with the
increase of the number of vector components. Classically, the most effective
algorithm, fast Fourier transform is in time $O(NlogN)$. On the contrary,
the quantum Fourier transform can be defined as a unitary transformation on $%
n$ qubits \cite{Nielsen}, which is:%
\begin{equation}
\hat{F}\left\vert j\right\rangle =\frac{1}{\sqrt{2^{n}}}\sum%
\limits_{k=0}^{2^{n}-1}e^{\frac{2\pi i}{2^{n}}jk}\left\vert k\right\rangle .
\label{eq2}
\end{equation}%
Furthermore, the quantum Fourier transform of arbitrary state $\left\vert
\Psi \right\rangle =C_{0}\left\vert 0\right\rangle +\cdots
+C_{2^{n}-1}\left\vert 2^{n}-1\right\rangle $ can be expressed as:%
\begin{eqnarray}
\left\vert \Psi \right\rangle _{F} &\equiv &\hat{F}\left\vert \Psi
\right\rangle =C_{0}\hat{F}\left\vert 0\right\rangle +C_{1}\hat{F}\left\vert
1\right\rangle +\cdots +C_{2^{n}-1}\hat{F}\left\vert 2^{n}-1\right\rangle
\label{eq3} \\
&=&\frac{1}{\sqrt{2^{n}}}\sum\limits_{k=0}^{2^{n}-1}\left[ C_{0}\omega
^{0\ast k}+C_{1}\omega ^{1\ast k}+\cdots +C_{2^{n}-1}\omega ^{\left(
2^{n}-1\right) \ast k}\right] \left\vert k\right\rangle ,  \nonumber
\end{eqnarray}%
where $\omega =e^{2\pi i/2^{n}}$. Then, we expand $\left\vert \Psi
\right\rangle _{F}$ into:%
\begin{equation}
\left\vert \Psi \right\rangle _{F}=\sum\limits_{j_{1}=0}^{1}\cdots
\sum\limits_{j_{n}=0}^{1}D_{j_{n-1}\cdots j_{0}}\left\vert j_{n-1}\cdots
j_{0}\right\rangle ,  \label{eq4}
\end{equation}%
where the coefficients satisfy the following equation:%
\begin{equation}
\left( 
\begin{array}{c}
D_{0} \\ 
D_{1} \\ 
\vdots \\ 
D_{2^{n}-1}%
\end{array}%
\right) =\frac{1}{\sqrt{2^{n}}}\left( 
\begin{array}{cccc}
1 & 1 & \cdots & 1 \\ 
1 & \omega & \cdots & \omega ^{2^{n}-1} \\ 
\vdots & \vdots & \ddots & \vdots \\ 
1 & \omega ^{2^{n}-1} & \cdots & \omega ^{\left( 2^{n}-1\right) ^{2}}%
\end{array}%
\right) \left( 
\begin{array}{c}
C_{0} \\ 
C_{1} \\ 
\vdots \\ 
C_{2^{n}-1}%
\end{array}%
\right) .  \label{eq5}
\end{equation}%
In quantum Fourier transform, after the Hadamard gate and controlled-phase
gate, we can obtain the final state $\left\vert j\right\rangle =\left\vert
j_{n-1}j_{n-2}\cdots j_{0}\right\rangle $ of quantum Fourier transform:%
\begin{equation}
\hat{F}\left\vert j\right\rangle =\frac{1}{\sqrt{2^{n}}}\left( \left\vert
1\right\rangle +e^{2\pi i0.j_{0}}\left\vert 1\right\rangle \right) \left(
\left\vert 1\right\rangle +e^{2\pi i0.j_{1}j_{0}}\left\vert 1\right\rangle
\right) \cdots \left( \left\vert 1\right\rangle +e^{2\pi
i0.j_{n-1}j_{n-2}\cdots j_{0}}\left\vert 1\right\rangle \right) .
\label{eq6}
\end{equation}%
There are n Hadamard gates and $n(n-1)/2$ controlled-phase gates on n qubit
registers, which means the quantum Fourier transform takes $O(n^{2})$ basic
gate operations. Nevertheless, the quantum Fourier transform cannot output
precise result of final states directly, but the probability of every state
by repeated measurements, which can output the final result of Fourier
transform at a certain accuracy \cite{Nielsen}.

\section{Optical analogy to quantum Fourier transform \label{sec3}}

\subsection{Simulation of quantum states based on pseudorandom phase
ensemble \label{sec3.1}}

In Ref. \cite{Fu1,Fu2,Fu3}, a way to simulation of quantum states was
introduced, which utilize the properties of pseudorandom sequence to
modulate classical optical fields into different quantum states with
different pseudorandom sequences. The formal product states for these fields
can be a simulation to arbitrary quantum states in the pseudorandom phase
ensemble model \cite{Fu3}. To further illustrate this method, the following
is a brief introduction.

There are two orthogonal modes (polarization or transverse) of a classical
field, which are denoted by $\left\vert 0\right\rangle $ and $\left\vert
1\right\rangle $, respectively. Thus, a qubit state $\left\vert \psi
\right\rangle =\alpha \left\vert 0\right\rangle +\beta \left\vert
1\right\rangle $ can be expressed by the mode superposition, where $|\alpha
|^{2}+|\beta |^{2}=1,(\alpha ,\beta \in C)$. Obviously, all the mode
superposition states span a Hilbert space. Choosing any $n$ PPSs from the
set $\Xi =\{\lambda ^{(0)},\lambda ^{(1)},\ldots \lambda ^{(M-1)}\}$ over $%
GF(p)$ to modulate $n$ classical fields, we can obtain the states expressed
as follows:%
\begin{equation}
\begin{array}{c}
\left\vert \psi _{1}\right\rangle =e^{i\lambda ^{(1)}}\left( \alpha
_{1}\left\vert 0\right\rangle +\beta _{1}\left\vert 1\right\rangle \right) ,
\\ 
\vdots \\ 
\left\vert \psi _{n}\right\rangle =e^{i\lambda ^{(n)}}\left( \alpha
_{n}\left\vert 0\right\rangle +\beta _{n}\left\vert 1\right\rangle \right) .%
\end{array}
\label{eq7}
\end{equation}

According to the properties of PPSs and Hilbert space, we can define the
inner product of any two fields $\left\vert \psi _{a}\right\rangle $ and $%
\left\vert \psi _{b}\right\rangle $. We obtain the orthogonal property in
our simulation,%
\begin{equation}
\left\langle \psi _{a}\left\vert \psi _{b}\right. \right\rangle =\frac{1}{M}%
\sum\limits_{k=1}^{M}e^{i(\lambda _{k}^{(b)}-\lambda _{k}^{(a)})}(\alpha
_{a}^{\ast }\alpha _{b}+\beta _{a}^{\ast }\beta _{b})=\left\{ 
\begin{array}{c}
1,a=b, \\ 
0,a\neq b,%
\end{array}%
\right.   \label{eq8}
\end{equation}%
where $\lambda _{k}^{(a)},\lambda _{k}^{(b)}$ are the $k$-th units of $%
\lambda ^{(a)}$ and $\lambda ^{(b)}$, respectively. The orthogonal property
supports the construction of the tensor product structure of the multiple
states. A formal product state $\left\vert \Psi \right\rangle $ for the $n$
classical fields is defined as being a direct product of $\left\vert \psi
_{i}\right\rangle ,$%
\begin{equation}
\left\vert \Psi \right\rangle =\left\vert \psi _{1}\right\rangle \otimes
\left\vert \psi _{2}\right\rangle \otimes \cdots \otimes \left\vert \psi
_{n}\right\rangle .  \label{eq9}
\end{equation}

According to the definition, $n$ classical fields of Eq. (\ref{eq7}) can be
expressed as the following states:%
\begin{equation}
\left\vert \Psi \right\rangle =e^{i\sum\nolimits_{j=1}^{n}\lambda ^{\left(
j\right) }}\left( \left\vert 0\right\rangle +\left\vert 1\right\rangle
+\cdots +\left\vert 2^{n}-1\right\rangle \right) .  \label{eq10}
\end{equation}

As mentioned in \cite{Fu1,Fu2}, a general form of $\left\vert \psi
_{k}\right\rangle $ for $n$ fields can be constructed from Eq. (\ref{eq7})
using a gate array model,%
\begin{eqnarray}
\left\vert \psi _{k}\right\rangle &=&\sum\limits_{i=1}^{n}\alpha
_{k}^{\left( i\right) }e^{i\lambda ^{\left( i\right) }}\left\vert
0\right\rangle +\sum\limits_{j=1}^{n}\beta _{k}^{\left( j\right)
}e^{i\lambda ^{\left( j\right) }}\left\vert 1\right\rangle  \label{eq11} \\
&\equiv &\tilde{\alpha}_{k}\left\vert 0\right\rangle +\tilde{\beta}%
_{k}\left\vert 1\right\rangle ,  \nonumber
\end{eqnarray}%
where $\tilde{\alpha}_{k}\equiv \sum\limits_{i=1}^{n}\alpha _{k}^{\left(
i\right) }e^{i\lambda ^{\left( i\right) }},\tilde{\beta}_{k}\equiv
\sum\limits_{j=1}^{n}\beta _{k}^{\left( j\right) }e^{i\lambda ^{\left(
j\right) }}$. Then, the formal product state (\ref{eq9}) can be written as%
\begin{equation}
\left\vert \Psi \right\rangle =\left( \tilde{\alpha}_{1}\left\vert
0\right\rangle +\tilde{\beta}_{1}\left\vert 1\right\rangle \right) \otimes
\cdots \otimes \left( \tilde{\alpha}_{n}\left\vert 0\right\rangle +\tilde{%
\beta}_{n}\left\vert 1\right\rangle \right) .  \label{eq12}
\end{equation}

Further, we can obtain each item of the superposition of $\left\vert \Psi
\right\rangle $ as follows:%
\begin{equation}
\begin{array}{c}
C_{00\cdots 0}\left\vert 00\cdots 0\right\rangle =\tilde{\alpha}_{1}\tilde{%
\alpha}_{2}\cdots \tilde{\alpha}_{n}\left\vert 00\cdots 0\right\rangle , \\ 
C_{00\cdots 1}\left\vert 00\cdots 1\right\rangle =\tilde{\alpha}_{1}\tilde{%
\alpha}_{2}\cdots \tilde{\beta}_{n}\left\vert 00\cdots 1\right\rangle , \\ 
\vdots \\ 
C_{11\cdots 1}\left\vert 11\cdots 1\right\rangle =\tilde{\beta}_{1}\tilde{%
\beta}_{2}\cdots \tilde{\beta}_{n}\left\vert 11\cdots 1\right\rangle .%
\end{array}
\label{eq13}
\end{equation}

According to the closure property \cite{Fu1}, the phase sequences $\lambda
^{\left( j\right) }$ of $C_{i_{1}i_{2}\cdots i_{n}}$ remain in the set $\Xi $%
, which means $C_{i_{1}i_{2}\cdots
i_{n}}=\sum\limits_{j=0}^{M-1}C_{i_{1}i_{2}\cdots i_{n}}^{\left( j\right)
}e^{i\lambda ^{\left( j\right) }}$. Therefore, we obtain the formal product
state $\left\vert \Psi \right\rangle $ spans a Hilbert space with the basis $%
\left\{ \left. e^{i\lambda ^{\left( j\right) }}\left\vert i_{1}i_{2}\cdots
i_{n}\right\rangle \right\vert \lambda ^{\left( j\right) }\in \Xi ,j=0\cdots
M-1,i_{n}=0or1\right\} $ and can be expressed as follows:%
\begin{equation}
\left\vert \Psi \right\rangle =\sum\limits_{i_{1}=0}^{1}\cdots
\sum\limits_{i_{n}=0}^{1}\left[ \sum\limits_{j=0}^{M-1}C_{i_{1}i_{2}\cdots
i_{n}}^{\left( j\right) }e^{i\lambda ^{\left( j\right) }}\left\vert
i_{1}i_{2}\cdots i_{n}\right\rangle \right] ,  \label{eq14}
\end{equation}%
where $C_{i_{1}i_{2}\cdots i_{n}}^{\left( j\right) }$ denotes a total of $%
M2^{n}$ coefficients.

Ref. \cite{Fu3} further proposed the ensemble-averaged reduced states in
ensemble model, which utilize the closure and balance property of
pseudorandom phase sequence to remove some terms in the formal product
states after ensemble averaging, whereby the required arbitrary quantum
states can be obtained, including quantum entanglement states.

\subsection{Algorithm for the optical analogy \label{sec3.2}}

Considering a quantum state $\left\vert \Psi \right\rangle $ expressed by $n$
classical fields, according to the equation (\ref{eq14}), we obtain its
formal product states as following:%
\begin{equation}
\left\vert \Psi \right\rangle =\sum\limits_{i_{1}=0}^{1}\cdots
\sum\limits_{i_{n}=0}^{1}C_{i_{1}i_{2}\cdots i_{n}}\left\vert
i_{1}i_{2}\cdots i_{n}\right\rangle ,  \label{eq15}
\end{equation}%
where $C_{i_{1}i_{2}\cdots i_{N}}$ is the same as that in Eq. (\ref{eq14}).
After quantum Fourier transform, this state involves into 
\begin{equation}
\left\vert \Psi \right\rangle _{F}=\hat{F}\left\vert \Psi \right\rangle ,
\label{eq16}
\end{equation}%
where 
\begin{equation}
\left\vert \Psi \right\rangle _{F}=\sum\limits_{j_{1}=0}^{1}\cdots
\sum\limits_{j_{n}=0}^{1}D_{j_{1}j_{2}\cdots j_{n}}\left\vert
j_{1}j_{2}\cdots j_{n}\right\rangle .  \label{eq17}
\end{equation}

According\ to the definition Eq. (\ref{eq1}), the relation between the
coefficients $C_{i_{1}i_{2}\cdots i_{n}}$ and $D_{j_{1}j_{2}\cdots j_{n}}$
of these two states have to be satisfied as Eq. (\ref{eq5}). To obtain the
relation between these coefficient, we design the following algortithm:

(1) Selected a basis state $\left\vert j_{1}j_{2}\cdots j_{n}\right\rangle $
of $\left\vert \Psi \right\rangle _{F}$;

(2) Apply the following controlled-phase transformation on every field of $%
\left\vert \Psi \right\rangle $ according to the specific value of bits in
the selected basis state:%
\begin{equation}
\left\{ 
\begin{array}{c}
\left\vert \psi _{1}\right\rangle =\tilde{\alpha}_{1}\left\vert
0\right\rangle +\tilde{\beta}_{1}\left\vert 1\right\rangle \\ 
\left\vert \psi _{2}\right\rangle =\tilde{\alpha}_{2}\left\vert
0\right\rangle +\omega ^{j_{1}\ast 2^{n-2}}\tilde{\beta}_{2}\left\vert
1\right\rangle \\ 
\left\vert \psi _{3}\right\rangle =\tilde{\alpha}_{3}\left\vert
0\right\rangle +\omega ^{j_{2}\ast 2^{n-2}+j_{1}\ast 2^{n-3}}\tilde{\beta}%
_{3}\left\vert 1\right\rangle \\ 
\cdots \\ 
\left\vert \psi _{n}\right\rangle =\tilde{\alpha}_{n}\left\vert
0\right\rangle +\omega ^{j_{n-1}\ast 2^{n-2}+j_{n-2}\ast 2^{n-3}+\cdots
+j_{1}\ast 1}\tilde{\beta}_{n}\left\vert 1\right\rangle%
\end{array}%
\right.  \label{eq18}
\end{equation}

(3) Apply Hadamard gate on these fields, after which we obtain:%
\begin{equation}
\left\{ 
\begin{array}{c}
\left\vert \psi _{1}\right\rangle =\left( \tilde{\alpha}_{1}+\tilde{\beta}%
_{1}\right) \left\vert 0\right\rangle +\left( \tilde{\alpha}_{1}-\tilde{\beta%
}_{1}\right) \left\vert 1\right\rangle \\ 
\left\vert \psi _{2}\right\rangle =\left( \tilde{\alpha}_{2}+\omega
^{j_{1}\ast 2^{n-2}}\tilde{\beta}_{2}\right) \left\vert 0\right\rangle
+\left( \tilde{\alpha}_{2}-\omega ^{j_{1}\ast 2^{n-2}}\tilde{\beta}%
_{2}\right) \left\vert 1\right\rangle \\ 
\left\vert \psi _{3}\right\rangle =\left( \tilde{\alpha}_{3}+\omega
^{j_{2}\ast 2^{n-2}+j_{1}\ast 2^{n-3}}\tilde{\beta}_{3}\right) \left\vert
0\right\rangle +\left( \tilde{\alpha}_{3}-\omega ^{j_{2}\ast
2^{n-2}+j_{1}\ast 2^{n-3}}\tilde{\beta}_{3}\right) \left\vert 1\right\rangle
\\ 
\cdots \\ 
\left\vert \psi _{n}\right\rangle =\left( \tilde{\alpha}_{n}+\omega
^{j_{n-1}\ast 2^{n-2}+j_{n-2}\ast 2^{n-3}+\cdots +j_{1}}\tilde{\beta}%
_{n}\right) \tilde{\alpha}_{n}\left\vert 0\right\rangle +\left( \tilde{\alpha%
}_{n}-\omega ^{j_{n-1}\ast 2^{n-2}+j_{n-2}\ast 2^{n-3}+\cdots +j_{1}}\tilde{%
\beta}_{n}\right) \left\vert 1\right\rangle%
\end{array}%
\right.  \label{eq19}
\end{equation}

(4) Apply the mode selection gate on these fields according to the specific
values in $\left\vert j_{1}j_{2}\cdots j_{n}\right\rangle $, after which the
mode of every field is identical to the corresponding value in $\left\vert
j_{1}j_{2}\cdots j_{n}\right\rangle $, e.g., if $j_{1}=0$, $\left\vert \psi
_{1}\right\rangle \rightarrow \left( \tilde{\alpha}_{1}+\tilde{\beta}%
_{1}\right) \left\vert 0\right\rangle $, while if $j_{1}=1$, $\left\vert
\psi _{1}\right\rangle \rightarrow \left( \tilde{\alpha}_{1}-\tilde{\beta}%
_{1}\right) \left\vert 1\right\rangle $;

(5) Apply mode detection on these fields and obtain the M matrix. Then we
can obtain the corresponding coefficient $D_{j_{n}j_{n-1}\cdots j_{1}}$
using the method in Ref. \cite{Fu2}.

\begin{figure}[htbp]
\centering\includegraphics[height=7.3609cm,width=13.2369cm]{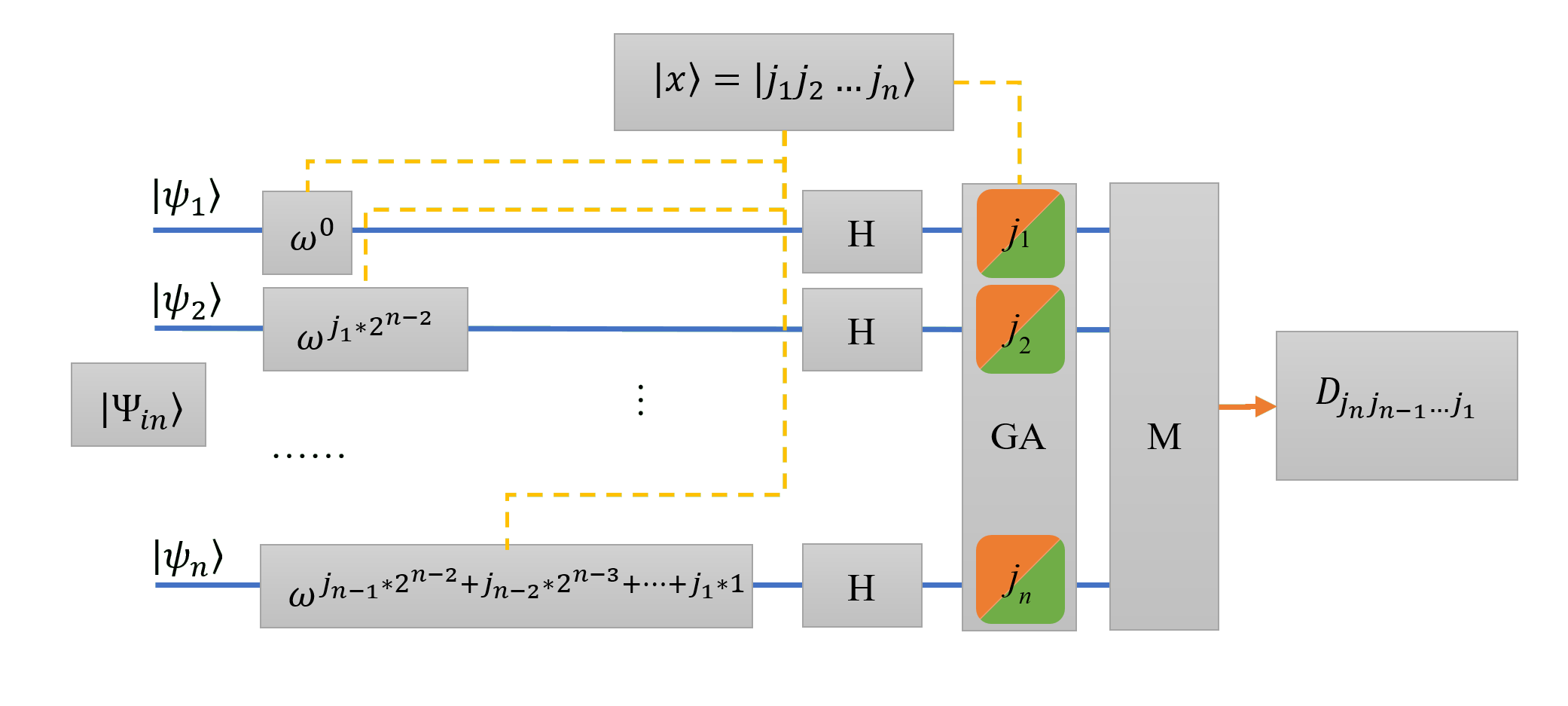}
\caption{The algorithm diagram of optical analogy to quantum Fourier
transformation.}
\label{fig1}
\end{figure}

The above algorithm can be summarized as the following block diagram in Fig. %
\ref{fig1}.

At last, we can analysis the computational complexity: there are $n$ fields
in $\left\vert \Psi \right\rangle $ after $n$ controlled-phase gates, $n$
Hadamard gates, $n$ mode selection operations and finally $n^{2}$
correlation detection in mode detection. Hence, the total number of
operations is in $O(n^{2})$, which is the same as that in quantum Fourier
transform. However, the result we obtain is with certainty but not with
probability like the case in quantum Fourier transform.

\subsection{The equivalence of ensemble-averaged reduced states in optical
analogy algorithm \label{sec3.3}}

In pseudorandom phase ensemble, we utilize the characteristic of
pseudorandom phase sequence to define the ensemble-averaged reduced state 
\cite{Fu3}. The balance property of pseudorandom phase sequences \cite{Fu1}:
with the exception of $\lambda ^{\left( 0\right) }$, any sequence of the set 
$\Omega $ satisfies%
\begin{equation}
\sum\limits_{k=1}^{M}e^{i\theta }e^{i\lambda _{k}^{\left( j\right)
}}=\sum\limits_{k=1}^{M}e^{i\left( \theta +\lambda _{k}^{\left( j\right)
}\right) }=0,\forall \theta \in 
%TCIMACRO{\U{211d} }%
%BeginExpansion
\mathbb{R}
%EndExpansion
\label{eq20}
\end{equation}%
where $\lambda _{k}^{\left( j\right) }$ is the $k$-th phase unit in
pseudorandom sequence $\lambda ^{\left( j\right) }$. Due to this property,
several terms of the formal product state can be reduced under ensemble
averaging, enabling us to utilize classical fields to simulation arbitrary
quantum states. According to Ref. \cite{Fu3}, the reduced state can be
defined as%
\begin{equation}
\left\vert \tilde{\Psi}\right\rangle \equiv
\sum\limits_{k=1}^{M}e^{-i\lambda ^{\left( S\right) }}\left\vert \Psi
\right\rangle  \label{eq21}
\end{equation}%
where $\lambda ^{\left( S\right) }=\sum\nolimits_{k=1}^{n}\lambda ^{(k)}$,
which is the sum of all phase sequences of classical fields.

Then, we discuss about the quantum Fourier transform of ensemble-averaged
reduced states. From equation (\ref{eq5}), we obtain the coefficients of
quantum Fourier transform satisfies%
\begin{equation}
\begin{array}{c}
D_{00\cdots 0}=C_{00\cdots 0}+C_{00\cdots 1}+\cdots +C_{11\cdots 1} \\ 
D_{00\cdots 1}=C_{00\cdots 0}+\omega C_{00\cdots 1}+\cdots +\omega ^{\left(
2^{n}-1\right) }C_{11\cdots 1} \\ 
\cdots \\ 
D_{11\cdots 0}=C_{00\cdots 0}+\omega ^{\left( 2^{n}-2\right) }C_{00\cdots
1}+\cdots +\omega ^{\left( 2^{n}-2\right) \left( 2^{n}-1\right) }C_{11\cdots
1} \\ 
D_{11\cdots 1}=C_{00\cdots 0}+\omega ^{\left( 2^{n}-1\right) }C_{00\cdots
1}+\cdots +\omega ^{\left( 2^{n}-1\right) ^{2}}C_{11\cdots 1}%
\end{array}
\label{eq22}
\end{equation}

In these equations, the combinations of $\omega $ and $C_{i_{1}i_{2}\cdots
i_{n}}$ satisfy the following relations:%
\begin{equation}
\omega ^{k}C_{i_{1}i_{2}\cdots
i_{n}}=\sum\limits_{j=1}^{M}C_{i_{1}i_{2}\cdots i_{n}}^{\left( j\right)
}e^{i \left[ \lambda ^{\left( j\right) }+2\pi k/2^{n}\right] }  \label{eq23}
\end{equation}

Obviously, these terms also satisfy the balance property of pseudorandom
sequence. Hence, the ensemble-averaged reduced states can also be used in
the states after quantum Fourier transform. Then, we can obtain the Fourier
transform:%
\begin{equation}
\left\vert \tilde{\Psi}\right\rangle _{F}\equiv
\sum\limits_{k=1}^{M}e^{-i\lambda ^{\left( S\right) }}\left\vert \Psi
\right\rangle _{F}=\hat{F}\sum\limits_{k=1}^{M}e^{-i\lambda ^{\left(
S\right) }}\left\vert \Psi \right\rangle =\hat{F}\left\vert \tilde{\Psi}%
\right\rangle  \label{eq24}
\end{equation}

At last, we show the equivalence of ensemble-averaged reduced states in
quantum Fourier transform.

\section{Optical analogy to quantum Fourier transform for three particles 
\label{sec4}}

According to Ref. \cite{Fu2}, three pseudorandom phase sequences $\lambda
^{\left( i\right) }\left( i=1,2,3\right) $ are required to implement the
simulation of the quantum states consisting of three particles. Modulated
with these phase sequence, three classical optical fields can be expressed
as following:%
\begin{equation}
\left\{ 
\begin{array}{c}
\left\vert \psi _{1}\right\rangle =e^{i\lambda ^{\left( 1\right) }}\left(
\left\vert 0\right\rangle +\left\vert 1\right\rangle \right) \\ 
\left\vert \psi _{2}\right\rangle =e^{i\lambda ^{\left( 2\right) }}\left(
\left\vert 0\right\rangle +\left\vert 1\right\rangle \right) \\ 
\left\vert \psi _{3}\right\rangle =e^{i\lambda ^{\left( 3\right) }}\left(
\left\vert 0\right\rangle +\left\vert 1\right\rangle \right)%
\end{array}%
\right.  \label{eq25}
\end{equation}%
After proper gate array operation \cite{Fu3}, we can obtain arbitrary
quantum states which can be expressed as following:%
\begin{equation}
\left\{ 
\begin{array}{c}
\left\vert \psi _{1}\right\rangle =\tilde{\alpha}_{1}\left\vert
0\right\rangle +\tilde{\beta}_{1}\left\vert 1\right\rangle \\ 
\left\vert \psi _{2}\right\rangle =\tilde{\alpha}_{2}\left\vert
0\right\rangle +\tilde{\beta}_{2}\left\vert 1\right\rangle \\ 
\left\vert \psi _{3}\right\rangle =\tilde{\alpha}_{3}\left\vert
0\right\rangle +\tilde{\beta}_{3}\left\vert 1\right\rangle%
\end{array}%
\right.  \label{eq26}
\end{equation}

According to the algorithm in Sec. \ref{sec3.2}, we obtain:

(1) Apply controlled-phase gates on three classical fields respectively:

\begin{equation}
\left\{ 
\begin{array}{c}
\left\vert \psi _{1}\right\rangle =\tilde{\alpha}_{1}\left\vert
0\right\rangle +\tilde{\beta}_{1}\left\vert 1\right\rangle \\ 
\left\vert \psi _{2}\right\rangle =\tilde{\alpha}_{2}\left\vert
0\right\rangle +\omega ^{j_{1}\ast 2}\tilde{\beta}_{2}\left\vert
1\right\rangle \\ 
\left\vert \psi _{3}\right\rangle =\tilde{\alpha}_{3}\left\vert
0\right\rangle +\omega ^{j_{2}\ast 2+j_{1}\ast 1}\tilde{\beta}_{3}\left\vert
1\right\rangle%
\end{array}%
\right.  \label{eq27}
\end{equation}%
where $\omega =e^{2\pi i/8}$.

(2) Hadamard transformation%
\begin{equation}
\left\{ 
\begin{array}{c}
\left\vert \psi _{1}\right\rangle =\left( \tilde{\alpha}_{1}+\tilde{\beta}%
_{1}\right) \left\vert 0\right\rangle +\left( \tilde{\alpha}_{1}-\tilde{\beta%
}_{1}\right) \left\vert 1\right\rangle \\ 
\left\vert \psi _{2}\right\rangle =\left( \tilde{\alpha}_{2}+\omega
^{j_{1}\ast 2}\tilde{\beta}_{2}\right) \left\vert 0\right\rangle +\left( 
\tilde{\alpha}_{2}-\omega ^{j_{1}\ast 2}\tilde{\beta}_{2}\right) \left\vert
1\right\rangle \\ 
\left\vert \psi _{3}\right\rangle =\left( \tilde{\alpha}_{3}+\omega
^{j_{2}\ast 2+j_{1}\ast 1}\tilde{\beta}_{3}\right) \left\vert 0\right\rangle
+\left( \tilde{\alpha}_{3}-\omega ^{j_{2}\ast 2+j_{1}\ast 1}\tilde{\beta}%
_{3}\right) \left\vert 1\right\rangle%
\end{array}%
\right.  \label{eq28}
\end{equation}

(3) Calculate coefficients

(3.1) When $\left\vert j_{1}j_{2}j_{3}\right\rangle =\left\vert
000\right\rangle $ and $\left\vert j_{1}j_{2}j_{3}\right\rangle =\left\vert
001\right\rangle $,%
\begin{equation}
\left\{ 
\begin{array}{c}
\left\vert \psi _{1}\right\rangle =\tilde{\alpha}_{1}\left\vert
0\right\rangle +\tilde{\beta}_{1}\left\vert 1\right\rangle \\ 
\left\vert \psi _{2}\right\rangle =\tilde{\alpha}_{2}\left\vert
0\right\rangle +\omega ^{0\ast 2}\tilde{\beta}_{2}\left\vert 1\right\rangle
\\ 
\left\vert \psi _{3}\right\rangle =\tilde{\alpha}_{3}\left\vert
0\right\rangle +\omega ^{0\ast 2+0\ast 1}\tilde{\beta}_{3}\left\vert
1\right\rangle%
\end{array}%
\right. \rightarrow \left\{ 
\begin{array}{c}
\left\vert \psi _{1}\right\rangle =\left( \tilde{\alpha}_{1}+\tilde{\beta}%
_{1}\right) \left\vert 0\right\rangle +\left( \tilde{\alpha}_{1}-\tilde{\beta%
}_{1}\right) \left\vert 1\right\rangle \\ 
\left\vert \psi _{2}\right\rangle =\left( \tilde{\alpha}_{2}+\tilde{\beta}%
_{2}\right) \left\vert 0\right\rangle +\left( \tilde{\alpha}_{2}-\tilde{\beta%
}_{2}\right) \left\vert 1\right\rangle \\ 
\left\vert \psi _{3}\right\rangle =\left( \tilde{\alpha}_{3}+\tilde{\beta}%
_{3}\right) \left\vert 0\right\rangle +\left( \tilde{\alpha}_{3}-\tilde{\beta%
}_{3}\right) \left\vert 1\right\rangle%
\end{array}%
\right. .  \label{eq29}
\end{equation}%
Then obtain the corresponding coefficients $D_{000}$ and $D_{100}$:%
\begin{eqnarray}
D_{000} &=&\left( \tilde{\alpha}_{1}+\tilde{\beta}_{1}\right) \left( \tilde{%
\alpha}_{2}+\tilde{\beta}_{2}\right) \left( \tilde{\alpha}_{3}+\tilde{\beta}%
_{3}\right)  \label{eq30} \\
&=&C_{000}+C_{001}+C_{010}+C_{011}+C_{100}+C_{101}+C_{110}+C_{111}, 
\nonumber
\end{eqnarray}%
\begin{eqnarray}
D_{100} &=&\left( \tilde{\alpha}_{1}+\tilde{\beta}_{1}\right) \left( \tilde{%
\alpha}_{2}+\tilde{\beta}_{2}\right) \left( \tilde{\alpha}_{3}-\tilde{\beta}%
_{3}\right)  \label{eq31} \\
&=&C_{000}-C_{001}+C_{010}-C_{011}+C_{100}-C_{101}+C_{110}-C_{111}. 
\nonumber
\end{eqnarray}

(3.2) When $\left\vert j_{1}j_{2}j_{3}\right\rangle =\left\vert
010\right\rangle $ and $\left\vert j_{1}j_{2}j_{3}\right\rangle =\left\vert
011\right\rangle $,%
\begin{equation}
\left\{ 
\begin{array}{c}
\left\vert \psi _{1}\right\rangle =\tilde{\alpha}_{1}\left\vert
0\right\rangle +\tilde{\beta}_{1}\left\vert 1\right\rangle \\ 
\left\vert \psi _{2}\right\rangle =\tilde{\alpha}_{2}\left\vert
0\right\rangle +\omega ^{0\ast 2}\tilde{\beta}_{2}\left\vert 1\right\rangle
\\ 
\left\vert \psi _{3}\right\rangle =\tilde{\alpha}_{3}\left\vert
0\right\rangle +\omega ^{1\ast 2+0\ast 1}\tilde{\beta}_{3}\left\vert
1\right\rangle%
\end{array}%
\right. \rightarrow \left\{ 
\begin{array}{c}
\left\vert \psi _{1}\right\rangle =\left( \tilde{\alpha}_{1}+\tilde{\beta}%
_{1}\right) \left\vert 0\right\rangle +\left( \tilde{\alpha}_{1}-\tilde{\beta%
}_{1}\right) \left\vert 1\right\rangle \\ 
\left\vert \psi _{2}\right\rangle =\left( \tilde{\alpha}_{2}+\tilde{\beta}%
_{2}\right) \left\vert 0\right\rangle +\left( \tilde{\alpha}_{2}-\tilde{\beta%
}_{2}\right) \left\vert 1\right\rangle \\ 
\left\vert \psi _{3}\right\rangle =\left( \tilde{\alpha}_{3}+\omega ^{2}%
\tilde{\beta}_{3}\right) \left\vert 0\right\rangle +\left( \tilde{\alpha}%
_{3}-\omega ^{2}\tilde{\beta}_{3}\right) \left\vert 1\right\rangle%
\end{array}%
\right. .  \label{eq32}
\end{equation}%
Then obtain the corresponding coefficients $D_{010}$ and $D_{110}$:%
\begin{eqnarray}
D_{010} &=&\left( \tilde{\alpha}_{1}+\tilde{\beta}_{1}\right) \left( \tilde{%
\alpha}_{2}-\tilde{\beta}_{2}\right) \left( \tilde{\alpha}_{3}+\omega ^{2}%
\tilde{\beta}_{3}\right)  \label{eq33} \\
&=&C_{000}+\omega ^{2}C_{001}-C_{010}-\omega ^{2}C_{011}+C_{100}+\omega
^{2}C_{101}-C_{110}-\omega ^{2}C_{111},  \nonumber
\end{eqnarray}%
\begin{eqnarray}
D_{110} &=&\left( \tilde{\alpha}_{1}+\tilde{\beta}_{1}\right) \left( \tilde{%
\alpha}_{2}-\tilde{\beta}_{2}\right) \left( \tilde{\alpha}_{3}-\omega ^{2}%
\tilde{\beta}_{3}\right)  \label{eq34} \\
&=&C_{000}-\omega ^{2}C_{001}-C_{010}+\omega ^{2}C_{011}+C_{100}-\omega
^{2}C_{101}-C_{110}+\omega ^{2}C_{111}.  \nonumber
\end{eqnarray}

(3.3) When $\left\vert j_{1}j_{2}j_{3}\right\rangle =\left\vert
100\right\rangle $ and $\left\vert j_{1}j_{2}j_{3}\right\rangle =\left\vert
101\right\rangle $,%
\begin{equation}
\left\{ 
\begin{array}{c}
\left\vert \psi _{1}\right\rangle =\tilde{\alpha}_{1}\left\vert
0\right\rangle +\tilde{\beta}_{1}\left\vert 1\right\rangle \\ 
\left\vert \psi _{2}\right\rangle =\tilde{\alpha}_{2}\left\vert
0\right\rangle +\omega ^{1\ast 2}\tilde{\beta}_{2}\left\vert 1\right\rangle
\\ 
\left\vert \psi _{3}\right\rangle =\tilde{\alpha}_{3}\left\vert
0\right\rangle +\omega ^{0\ast 2+1\ast 1}\tilde{\beta}_{3}\left\vert
1\right\rangle%
\end{array}%
\right. \rightarrow \left\{ 
\begin{array}{c}
\left\vert \psi _{1}\right\rangle =\left( \tilde{\alpha}_{1}+\tilde{\beta}%
_{1}\right) \left\vert 0\right\rangle +\left( \tilde{\alpha}_{1}-\tilde{\beta%
}_{1}\right) \left\vert 1\right\rangle \\ 
\left\vert \psi _{2}\right\rangle =\left( \tilde{\alpha}_{2}+\omega ^{2}%
\tilde{\beta}_{2}\right) \left\vert 0\right\rangle +\left( \tilde{\alpha}%
_{2}-\omega ^{2}\tilde{\beta}_{2}\right) \left\vert 1\right\rangle \\ 
\left\vert \psi _{3}\right\rangle =\left( \tilde{\alpha}_{3}+\omega \tilde{%
\beta}_{3}\right) \left\vert 0\right\rangle +\left( \tilde{\alpha}%
_{3}-\omega \tilde{\beta}_{3}\right) \left\vert 1\right\rangle%
\end{array}%
\right. .  \label{eq35}
\end{equation}%
Then obtain the corresponding coefficients $D_{001}$ and $D_{101}$:%
\begin{eqnarray}
D_{001} &=&\left( \tilde{\alpha}_{1}-\tilde{\beta}_{1}\right) \left( \tilde{%
\alpha}_{2}+\omega ^{2}\tilde{\beta}_{2}\right) \left( \tilde{\alpha}%
_{3}+\omega \tilde{\beta}_{3}\right)  \label{eq36} \\
&=&C_{000}+\omega C_{001}+\omega ^{2}C_{010}+\omega
^{3}C_{011}-C_{100}-\omega C_{101}-\omega ^{2}C_{110}-\omega ^{3}C_{111}, 
\nonumber
\end{eqnarray}%
\begin{eqnarray}
D_{101} &=&\left( \tilde{\alpha}_{1}-\tilde{\beta}_{1}\right) \left( \tilde{%
\alpha}_{2}+\omega ^{2}\tilde{\beta}_{2}\right) \left( \tilde{\alpha}%
_{3}-\omega \tilde{\beta}_{3}\right)  \label{eq37} \\
&=&C_{000}-\omega C_{001}+\omega ^{2}C_{010}-\omega
^{3}C_{011}-C_{100}+\omega C_{101}-\omega ^{2}C_{110}+\omega ^{3}C_{111}. 
\nonumber
\end{eqnarray}

(3.4) When $\left\vert j_{1}j_{2}j_{3}\right\rangle =\left\vert
110\right\rangle $ and $\left\vert j_{1}j_{2}j_{3}\right\rangle =\left\vert
111\right\rangle $,%
\begin{equation}
\left\{ 
\begin{array}{c}
\left\vert \psi _{1}\right\rangle =\tilde{\alpha}_{1}\left\vert
0\right\rangle +\tilde{\beta}_{1}\left\vert 1\right\rangle \\ 
\left\vert \psi _{2}\right\rangle =\tilde{\alpha}_{2}\left\vert
0\right\rangle +\omega ^{1\ast 2}\tilde{\beta}_{2}\left\vert 1\right\rangle
\\ 
\left\vert \psi _{3}\right\rangle =\tilde{\alpha}_{3}\left\vert
0\right\rangle +\omega ^{1\ast 2+1\ast 1}\tilde{\beta}_{3}\left\vert
1\right\rangle%
\end{array}%
\right. \rightarrow \left\{ 
\begin{array}{c}
\left\vert \psi _{1}\right\rangle =\left( \tilde{\alpha}_{1}+\tilde{\beta}%
_{1}\right) \left\vert 0\right\rangle +\left( \tilde{\alpha}_{1}-\tilde{\beta%
}_{1}\right) \left\vert 1\right\rangle \\ 
\left\vert \psi _{2}\right\rangle =\left( \tilde{\alpha}_{2}+\omega ^{2}%
\tilde{\beta}_{2}\right) \left\vert 0\right\rangle +\left( \tilde{\alpha}%
_{2}-\omega ^{2}\tilde{\beta}_{2}\right) \left\vert 1\right\rangle \\ 
\left\vert \psi _{3}\right\rangle =\left( \tilde{\alpha}_{3}+\omega ^{3}%
\tilde{\beta}_{3}\right) \left\vert 0\right\rangle +\left( \tilde{\alpha}%
_{3}-\omega ^{3}\tilde{\beta}_{3}\right) \left\vert 1\right\rangle%
\end{array}%
\right. .  \label{eq38}
\end{equation}%
Then obtain the corresponding coefficients $D_{011}$ and $D_{111}$:%
\begin{eqnarray}
D_{011} &=&\left( \tilde{\alpha}_{1}-\tilde{\beta}_{1}\right) \left( \tilde{%
\alpha}_{2}-\omega ^{2}\tilde{\beta}_{2}\right) \left( \tilde{\alpha}%
_{3}+\omega ^{3}\tilde{\beta}_{3}\right)  \label{eq39} \\
&=&C_{000}+\omega ^{3}C_{001}-\omega ^{2}C_{010}-\omega
^{5}C_{011}-C_{100}-\omega ^{3}C_{101}+\omega ^{2}C_{110}+\omega ^{5}C_{111},
\nonumber
\end{eqnarray}%
\begin{eqnarray}
D_{110} &=&\left( \tilde{\alpha}_{1}-\tilde{\beta}_{1}\right) \left( \tilde{%
\alpha}_{2}-\tilde{\beta}_{2}\right) \left( \tilde{\alpha}_{3}-\omega ^{2}%
\tilde{\beta}_{3}\right)  \label{eq40} \\
&=&C_{000}-\omega ^{3}C_{001}-\omega ^{2}C_{010}+\omega
^{5}C_{011}-C_{100}+\omega ^{3}C_{101}+\omega ^{2}C_{110}-\omega ^{5}C_{111}.
\nonumber
\end{eqnarray}%
At last, we obtain the transform matrix of all coefficients as following:%
\begin{equation}
\left( 
\begin{array}{c}
D_{000} \\ 
D_{001} \\ 
D_{010} \\ 
D_{011} \\ 
D_{100} \\ 
D_{101} \\ 
D_{110} \\ 
D_{111}%
\end{array}%
\right) =\left( 
\begin{array}{cccccccc}
1 & 1 & 1 & 1 & 1 & 1 & 1 & 1 \\ 
1 & \omega & \omega ^{2} & \omega ^{3} & -1 & -\omega & -\omega ^{2} & 
-\omega ^{3} \\ 
1 & \omega ^{2} & -1 & -\omega ^{2} & 1 & \omega ^{2} & -1 & \omega ^{2} \\ 
1 & \omega ^{3} & -\omega ^{2} & \omega & -1 & -\omega ^{3} & \omega ^{2} & 
-\omega \\ 
1 & -1 & 1 & -1 & 1 & -1 & 1 & -1 \\ 
1 & -\omega & \omega ^{2} & -\omega ^{3} & -1 & \omega & -\omega ^{2} & 
\omega ^{3} \\ 
1 & -\omega ^{2} & -1 & \omega ^{2} & 1 & -\omega ^{2} & -1 & \omega ^{2} \\ 
1 & -\omega ^{3} & -\omega ^{2} & -\omega & -1 & \omega ^{3} & \omega ^{2} & 
\omega%
\end{array}%
\right) \left( 
\begin{array}{c}
C_{000} \\ 
C_{001} \\ 
C_{010} \\ 
C_{011} \\ 
C_{100} \\ 
C_{101} \\ 
C_{110} \\ 
C_{111}%
\end{array}%
\right)  \label{eq41}
\end{equation}%
We will utilize the above algorithm to apply quantum Fourier transform on
several kinds of states in the following:

(1) Product state

In quantum mechanics, the product state of three particles is $\left\vert
\Psi \right\rangle =\frac{1}{\sqrt{8}}\left( \left\vert 000\right\rangle
+\left\vert 001\right\rangle +\cdots +\left\vert 111\right\rangle \right) $.
We can expressed these three fields as equation (\ref{eq25}), except for
normalization constant. In Ref. [6], the formal product state of this state
can be expressed as:%
\begin{equation}
\left\vert \Psi \right\rangle =\left\vert \psi _{1}\right\rangle \otimes
\left\vert \psi _{2}\right\rangle \otimes \left\vert \psi _{3}\right\rangle
=e^{i\left( \lambda ^{\left( 1\right) }+\lambda ^{\left( 2\right) }+\lambda
^{\left( 3\right) }\right) }\left( \left\vert 000\right\rangle +\left\vert
001\right\rangle +\cdots +\left\vert 111\right\rangle \right)  \label{eq42}
\end{equation}%
Except for normalization constant and overall phase factor, that is, the sum
of three pseudorandom sequence, there is not difference between this state
and the product state of three particles. From Ref. \cite{Fu2,Fu3},
utilizing the concept of pseudorandom phase ensemble and the properties of
pseudorandom sequence, we obtain the ensemble-averaged reduced state:%
\begin{equation}
\left\vert \tilde{\Psi}\right\rangle =\left\vert 000\right\rangle
+\left\vert 001\right\rangle +\cdots +\left\vert 111\right\rangle
\label{eq43}
\end{equation}

Using the above algorithm, we can easily obtain the coefficients of the
Fourier transform of this state, $D_{000}=C_{000}+C_{001}+\cdots
+C_{111}=8e^{i\left( \lambda ^{\left( 1\right) }+\lambda ^{\left( 2\right)
}+\lambda ^{\left( 3\right) }\right) }$, while the other terms is 0. Then we
obtain:%
\begin{equation}
\left\vert \tilde{\Psi}\right\rangle _{F}=8\left\vert 000\right\rangle
\label{eq44}
\end{equation}%
which is identical to the quantum Fourier transform, except for the
normalization constant.

(2) GHZ state

In quantum mechanics, GHZ state is biggest entanglement state in the system
of the three particles. This state is of great importance since it can
verify the entanglement criterion in the correlation measurement of quantum
entanglement. From Ref. \cite{Fu2}, we can obtain the following form of
three fields by proper transformation:%
\begin{equation}
\left\{ 
\begin{array}{c}
\left\vert \psi _{1}\right\rangle =\tilde{\alpha}_{1}\left\vert
0\right\rangle +\tilde{\beta}_{1}\left\vert 1\right\rangle =e^{i\lambda
^{\left( 1\right) }}\left\vert 0\right\rangle +e^{i\lambda ^{\left( 2\right)
}}\left\vert 1\right\rangle \\ 
\left\vert \psi _{2}\right\rangle =\tilde{\alpha}_{2}\left\vert
0\right\rangle +\tilde{\beta}_{2}\left\vert 1\right\rangle =e^{i\lambda
^{\left( 2\right) }}\left\vert 0\right\rangle +e^{i\lambda ^{\left( 3\right)
}}\left\vert 1\right\rangle \\ 
\left\vert \psi _{3}\right\rangle =\tilde{\alpha}_{3}\left\vert
0\right\rangle +\tilde{\beta}_{3}\left\vert 1\right\rangle =e^{i\lambda
^{\left( 3\right) }}\left\vert 0\right\rangle +e^{i\lambda ^{\left( 1\right)
}}\left\vert 1\right\rangle%
\end{array}%
\right.  \label{eq45}
\end{equation}%
The formal product state of these three fields can be expressed as:%
\begin{eqnarray}
\left\vert \Psi \right\rangle &=&\left\vert \psi _{1}\right\rangle \otimes
\left\vert \psi _{2}\right\rangle \otimes \left\vert \psi _{3}\right\rangle
=e^{i\left( \lambda ^{\left( 1\right) }+\lambda ^{\left( 2\right) }+\lambda
^{\left( 3\right) }\right) }\left[ \left\vert 000\right\rangle +\left\vert
111\right\rangle +e^{i\left( \lambda ^{\left( 1\right) }-\lambda ^{\left(
3\right) }\right) }\left\vert 001\right\rangle \right.  \label{eq46} \\
&&+e^{i\left( \lambda ^{\left( 3\right) }-\lambda ^{\left( 2\right) }\right)
}\left\vert 010\right\rangle +e^{i\left( \lambda ^{\left( 1\right) }-\lambda
^{\left( 2\right) }\right) }\left\vert 011\right\rangle +e^{i\left( \lambda
^{\left( 2\right) }-\lambda ^{\left( 1\right) }\right) }\left\vert
100\right\rangle  \nonumber \\
&&\left. +e^{i\left( \lambda ^{\left( 2\right) }-\lambda ^{\left( 3\right)
}\right) }\left\vert 101\right\rangle +e^{i\left( \lambda ^{\left( 3\right)
}-\lambda ^{\left( 1\right) }\right) }\left\vert 110\right\rangle \right] 
\nonumber
\end{eqnarray}%
From Ref. \cite{Fu2,Fu3}, utilizing the concept of pseudorandom phase
ensemble and the properties of pseudorandom sequence, we obtain the
ensemble-averaged reduced state:%
\begin{equation}
\left\vert \tilde{\Psi}\right\rangle =\left\vert 000\right\rangle
+\left\vert 111\right\rangle  \label{eq47}
\end{equation}%
Similarly, except for normalization constant and overall phase factor, the
state is identical to GHZ state.

Using the above algorithm, we can easily obtain the coefficients of the
Fourier transform of this state respectively:%
\begin{eqnarray}
D_{000} &=&\left( \tilde{\alpha}_{1}+\tilde{\beta}_{1}\right) \left( \tilde{%
\alpha}_{2}+\tilde{\beta}_{2}\right) \left( \tilde{\alpha}_{3}+\tilde{\beta}%
_{3}\right) =2e^{i\left( \lambda ^{\left( 1\right) }+\lambda ^{\left(
2\right) }+\lambda ^{\left( 3\right) }\right) }+e^{i\left( 2\lambda ^{\left(
1\right) }+\lambda ^{\left( 2\right) }\right) }+e^{i\left( 2\lambda ^{\left(
1\right) }+\lambda ^{\left( 3\right) }\right) }  \nonumber \\
&&+e^{i\left( 2\lambda ^{\left( 2\right) }+\lambda ^{\left( 1\right)
}\right) }+e^{i\left( 2\lambda ^{\left( 2\right) }+\lambda ^{\left( 3\right)
}\right) }+e^{i\left( 2\lambda ^{\left( 3\right) }+\lambda ^{\left( 1\right)
}\right) }+e^{i\left( 2\lambda ^{\left( 3\right) }+\lambda ^{\left( 2\right)
}\right) },  \label{eq48}
\end{eqnarray}%
\begin{eqnarray}
D_{001} &=&\left( \tilde{\alpha}_{1}-\tilde{\beta}_{1}\right) \left( \tilde{%
\alpha}_{2}+\omega ^{2}\tilde{\beta}_{2}\right) \left( \tilde{\alpha}%
_{3}+\omega \tilde{\beta}_{3}\right) =\left( 1-\omega ^{3}\right) e^{i\left(
\lambda ^{\left( 1\right) }+\lambda ^{\left( 2\right) }+\lambda ^{\left(
3\right) }\right) }+\omega e^{i\left( 2\lambda ^{\left( 1\right) }+\lambda
^{\left( 2\right) }\right) }  \nonumber \\
&&+\omega ^{3}e^{i\left( 2\lambda ^{\left( 1\right) }+\lambda ^{\left(
3\right) }\right) }-\omega e^{i\left( 2\lambda ^{\left( 2\right) }+\lambda
^{\left( 1\right) }\right) }-e^{i\left( 2\lambda ^{\left( 2\right) }+\lambda
^{\left( 3\right) }\right) }+\omega ^{2}e^{i\left( 2\lambda ^{\left(
3\right) }+\lambda ^{\left( 1\right) }\right) }  \nonumber \\
&&-\omega ^{2}e^{i\left( 2\lambda ^{\left( 3\right) }+\lambda ^{\left(
2\right) }\right) },  \label{eq49}
\end{eqnarray}%
\begin{eqnarray}
D_{010} &=&\left( \tilde{\alpha}_{1}+\tilde{\beta}_{1}\right) \left( \tilde{%
\alpha}_{2}-\tilde{\beta}_{2}\right) \left( \tilde{\alpha}_{3}+\omega ^{2}%
\tilde{\beta}_{3}\right) =\left( 1-\omega ^{2}\right) e^{i\left( \lambda
^{\left( 1\right) }+\lambda ^{\left( 2\right) }+\lambda ^{\left( 3\right)
}\right) }+\omega ^{2}e^{i\left( 2\lambda ^{\left( 1\right) }+\lambda
^{\left( 2\right) }\right) }  \nonumber \\
&&-\omega ^{2}e^{i\left( 2\lambda ^{\left( 1\right) }+\lambda ^{\left(
3\right) }\right) }+\omega ^{2}e^{i\left( 2\lambda ^{\left( 2\right)
}+\lambda ^{\left( 1\right) }\right) }+e^{i\left( 2\lambda ^{\left( 2\right)
}+\lambda ^{\left( 3\right) }\right) }-e^{i\left( 2\lambda ^{\left( 3\right)
}+\lambda ^{\left( 1\right) }\right) }  \nonumber \\
&&-e^{i\left( 2\lambda ^{\left( 3\right) }+\lambda ^{\left( 2\right)
}\right) },  \label{eq50}
\end{eqnarray}%
\begin{eqnarray}
D_{011} &=&\left( \tilde{\alpha}_{1}-\tilde{\beta}_{1}\right) \left( \tilde{%
\alpha}_{2}-\omega ^{2}\tilde{\beta}_{2}\right) \left( \tilde{\alpha}%
_{3}+\omega ^{3}\tilde{\beta}_{3}\right) =\left( 1+\omega ^{5}\right)
e^{i\left( \lambda ^{\left( 1\right) }+\lambda ^{\left( 2\right) }+\lambda
^{\left( 3\right) }\right) }+\omega ^{3}e^{i\left( 2\lambda ^{\left(
1\right) }+\lambda ^{\left( 2\right) }\right) }  \nonumber \\
&&-\omega ^{5}e^{i\left( 2\lambda ^{\left( 1\right) }+\lambda ^{\left(
3\right) }\right) }-\omega ^{3}e^{i\left( 2\lambda ^{\left( 2\right)
}+\lambda ^{\left( 1\right) }\right) }-e^{i\left( 2\lambda ^{\left( 2\right)
}+\lambda ^{\left( 3\right) }\right) }-\omega ^{2}e^{i\left( 2\lambda
^{\left( 3\right) }+\lambda ^{\left( 1\right) }\right) }  \nonumber \\
&&+\omega ^{2}e^{i\left( 2\lambda ^{\left( 3\right) }+\lambda ^{\left(
2\right) }\right) },  \label{eq51}
\end{eqnarray}%
\begin{eqnarray}
D_{100} &=&\left( \tilde{\alpha}_{1}+\tilde{\beta}_{1}\right) \left( \tilde{%
\alpha}_{2}+\tilde{\beta}_{2}\right) \left( \tilde{\alpha}_{3}-\tilde{\beta}%
_{3}\right) =-e^{i\left( 2\lambda ^{\left( 1\right) }+\lambda ^{\left(
2\right) }\right) }-e^{i\left( 2\lambda ^{\left( 1\right) }+\lambda ^{\left(
3\right) }\right) }  \nonumber \\
&&-e^{i\left( 2\lambda ^{\left( 2\right) }+\lambda ^{\left( 1\right)
}\right) }+e^{i\left( 2\lambda ^{\left( 2\right) }+\lambda ^{\left( 3\right)
}\right) }+e^{i\left( 2\lambda ^{\left( 3\right) }+\lambda ^{\left( 1\right)
}\right) }+e^{i\left( 2\lambda ^{\left( 3\right) }+\lambda ^{\left( 2\right)
}\right) },  \label{eq52}
\end{eqnarray}%
\begin{eqnarray}
D_{101} &=&\left( \tilde{\alpha}_{1}-\tilde{\beta}_{1}\right) \left( \tilde{%
\alpha}_{2}+\omega ^{2}\tilde{\beta}_{2}\right) \left( \tilde{\alpha}%
_{3}-\omega \tilde{\beta}_{3}\right) =\left( 1+\omega ^{3}\right) e^{i\left(
\lambda ^{\left( 1\right) }+\lambda ^{\left( 2\right) }+\lambda ^{\left(
3\right) }\right) }-\omega e^{i\left( 2\lambda ^{\left( 1\right) }+\lambda
^{\left( 2\right) }\right) }  \nonumber \\
&&-\omega ^{3}e^{i\left( 2\lambda ^{\left( 1\right) }+\lambda ^{\left(
3\right) }\right) }+\omega e^{i\left( 2\lambda ^{\left( 2\right) }+\lambda
^{\left( 1\right) }\right) }-e^{i\left( 2\lambda ^{\left( 2\right) }+\lambda
^{\left( 3\right) }\right) }+\omega ^{2}e^{i\left( 2\lambda ^{\left(
3\right) }+\lambda ^{\left( 1\right) }\right) }  \nonumber \\
&&-\omega ^{2}e^{i\left( 2\lambda ^{\left( 3\right) }+\lambda ^{\left(
2\right) }\right) },  \label{eq53}
\end{eqnarray}%
\begin{eqnarray}
D_{110} &=&\left( \tilde{\alpha}_{1}+\tilde{\beta}_{1}\right) \left( \tilde{%
\alpha}_{2}-\tilde{\beta}_{2}\right) \left( \tilde{\alpha}_{3}-\omega ^{2}%
\tilde{\beta}_{3}\right) =\left( 1+\omega ^{2}\right) e^{i\left( \lambda
^{\left( 1\right) }+\lambda ^{\left( 2\right) }+\lambda ^{\left( 3\right)
}\right) }-\omega ^{2}e^{i\left( 2\lambda ^{\left( 1\right) }+\lambda
^{\left( 2\right) }\right) }  \nonumber \\
&&+\omega ^{2}e^{i\left( 2\lambda ^{\left( 1\right) }+\lambda ^{\left(
3\right) }\right) }-\omega ^{2}e^{i\left( 2\lambda ^{\left( 2\right)
}+\lambda ^{\left( 1\right) }\right) }+e^{i\left( 2\lambda ^{\left( 2\right)
}+\lambda ^{\left( 3\right) }\right) }-e^{i\left( 2\lambda ^{\left( 3\right)
}+\lambda ^{\left( 1\right) }\right) }  \nonumber \\
&&-e^{i\left( 2\lambda ^{\left( 3\right) }+\lambda ^{\left( 2\right)
}\right) },  \label{eq54}
\end{eqnarray}%
\begin{eqnarray}
D_{111} &=&\left( \tilde{\alpha}_{1}-\tilde{\beta}_{1}\right) \left( \tilde{%
\alpha}_{2}-\omega ^{2}\tilde{\beta}_{2}\right) \left( \tilde{\alpha}%
_{3}-\omega ^{3}\tilde{\beta}_{3}\right) =\left( 1-\omega ^{5}\right)
e^{i\left( \lambda ^{\left( 1\right) }+\lambda ^{\left( 2\right) }+\lambda
^{\left( 3\right) }\right) }-\omega ^{3}e^{i\left( 2\lambda ^{\left(
1\right) }+\lambda ^{\left( 2\right) }\right) }  \nonumber \\
&&+\omega ^{5}e^{i\left( 2\lambda ^{\left( 1\right) }+\lambda ^{\left(
3\right) }\right) }+\omega ^{3}e^{i\left( 2\lambda ^{\left( 2\right)
}+\lambda ^{\left( 1\right) }\right) }-e^{i\left( 2\lambda ^{\left( 2\right)
}+\lambda ^{\left( 3\right) }\right) }-\omega ^{2}e^{i\left( 2\lambda
^{\left( 3\right) }+\lambda ^{\left( 1\right) }\right) }  \nonumber \\
&&+\omega ^{2}e^{i\left( 2\lambda ^{\left( 3\right) }+\lambda ^{\left(
2\right) }\right) },  \label{eq55}
\end{eqnarray}%
Utilizing ensemble-averaged reduced state, we obtain: 
\begin{eqnarray}
\left\vert \tilde{\Psi}\right\rangle _{F}
&=&\sum\limits_{k=1}^{M}e^{-i\lambda ^{\left( S\right) }}\left\vert \Psi
\right\rangle _{F}=2\left\vert 000\right\rangle +\left( 1-\omega ^{3}\right)
\left\vert 001\right\rangle +\left( 1-\omega ^{2}\right) \left\vert
010\right\rangle  \label{eq56} \\
&&+\left( 1-\omega \right) \left\vert 011\right\rangle +\left( 1+\omega
^{3}\right) \left\vert 101\right\rangle +\left( 1+\omega ^{2}\right)
\left\vert 110\right\rangle +\left( 1+\omega \right) \left\vert
111\right\rangle  \nonumber
\end{eqnarray}%
In conclusion, $\left\vert \tilde{\Psi}\right\rangle _{F}$ is the Fourier
transform of $\left\vert \tilde{\Psi}\right\rangle $ for GHZ states.

(3) W state

In quantum mechanics, W state is the most robust entanglement state $%
\left\vert \Psi \right\rangle =\frac{1}{\sqrt{3}}\left( \left\vert
100\right\rangle +\left\vert 010\right\rangle +\left\vert 001\right\rangle
\right) $. From Ref. \cite{Fu2}, by proper transformation on equation (\ref%
{eq25}), we can obtain the expression of these three fields as following:%
\begin{equation}
\left\{ 
\begin{array}{c}
\left\vert \psi _{1}\right\rangle =\tilde{\alpha}_{1}\left\vert
0\right\rangle +\tilde{\beta}_{1}\left\vert 1\right\rangle =e^{i\lambda
^{\left( 1\right) }}\left\vert 1\right\rangle +e^{i\lambda ^{\left( 2\right)
}}\left\vert 0\right\rangle +e^{i\lambda ^{\left( 3\right) }}\left\vert
0\right\rangle \\ 
\left\vert \psi _{2}\right\rangle =\tilde{\alpha}_{2}\left\vert
0\right\rangle +\tilde{\beta}_{2}\left\vert 1\right\rangle =e^{i\lambda
^{\left( 1\right) }}\left\vert 1\right\rangle +e^{i\lambda ^{\left( 2\right)
}}\left\vert 0\right\rangle +e^{i\lambda ^{\left( 3\right) }}\left\vert
0\right\rangle \\ 
\left\vert \psi _{3}\right\rangle =\tilde{\alpha}_{3}\left\vert
0\right\rangle +\tilde{\beta}_{3}\left\vert 1\right\rangle =e^{i\lambda
^{\left( 1\right) }}\left\vert 1\right\rangle +e^{i\lambda ^{\left( 2\right)
}}\left\vert 0\right\rangle +e^{i\lambda ^{\left( 3\right) }}\left\vert
0\right\rangle%
\end{array}%
\right.  \label{eq57}
\end{equation}%
The formal product state of these three fields can be expressed as:%
\begin{eqnarray}
\left\vert \Psi \right\rangle &=&\left\vert \psi _{1}\right\rangle \otimes
\left\vert \psi _{2}\right\rangle \otimes \left\vert \psi _{3}\right\rangle
=e^{i\left( \lambda ^{\left( 1\right) }+\lambda ^{\left( 2\right) }+\lambda
^{\left( 3\right) }\right) }\left\{ \left[ 1+e^{i\left( \lambda ^{\left(
2\right) }-\lambda ^{\left( 3\right) }\right) }+e^{i\left( \lambda ^{\left(
3\right) }-\lambda ^{\left( 2\right) }\right) }\right] \times \left(
\left\vert 100\right\rangle +\left\vert 010\right\rangle \right. \right. 
\nonumber \\
&&\left. +\left\vert 001\right\rangle \right) +\left[ e^{i\left( \lambda
^{\left( 1\right) }-\lambda ^{\left( 3\right) }\right) }+e^{i\left( \lambda
^{\left( 1\right) }-\lambda ^{\left( 2\right) }\right) }\right] \left(
\left\vert 011\right\rangle +\left\vert 110\right\rangle +\left\vert
101\right\rangle \right) +e^{i\left( 2\lambda ^{\left( 1\right) }-\lambda
^{\left( 2\right) }-\lambda ^{\left( 3\right) }\right) }\left\vert
111\right\rangle  \nonumber \\
&&\left. +2\left[ e^{i\left( 2\lambda ^{\left( 2\right) }-\lambda ^{\left(
1\right) }-\lambda ^{\left( 3\right) }\right) }+e^{i\left( 2\lambda ^{\left(
3\right) }-\lambda ^{\left( 2\right) }-\lambda ^{\left( 1\right) }\right)
}+e^{i\left( \lambda ^{\left( 2\right) }-\lambda ^{\left( 1\right) }\right)
}+e^{i\left( \lambda ^{\left( 3\right) }-\lambda ^{\left( 1\right) }\right) }%
\right] \left\vert 000\right\rangle \right\}  \label{eq58}
\end{eqnarray}%
From Ref. \cite{Fu2,Fu3}, utilizing the concept of pseudorandom phase
ensemble and the properties of pseudorandom sequence, we obtain the
ensemble-averaged reduced state:%
\begin{equation}
\left\vert \tilde{\Psi}\right\rangle =\left\vert 100\right\rangle
+\left\vert 010\right\rangle +\left\vert 001\right\rangle  \label{eq59}
\end{equation}%
Similarly, except for normalization constant and overall phase factor, the
state is identical to W state.

Using the above algorithm, we can easily obtain the coefficients of the
Fourier transform of this state respectively:%
\begin{eqnarray}
D_{000} &=&\left( \tilde{\alpha}_{1}+\tilde{\beta}_{1}\right) \left( \tilde{%
\alpha}_{2}+\tilde{\beta}_{2}\right) \left( \tilde{\alpha}_{3}+\tilde{\beta}%
_{3}\right) =6e^{i\left( \lambda ^{\left( 1\right) }+\lambda ^{\left(
2\right) }+\lambda ^{\left( 3\right) }\right) }+3\left[ e^{i\left( 2\lambda
^{\left( 1\right) }+\lambda ^{\left( 2\right) }\right) }\right.  \nonumber \\
&&\left. +e^{i\left( 2\lambda ^{\left( 1\right) }+\lambda ^{\left( 3\right)
}\right) }+e^{i\left( 2\lambda ^{\left( 2\right) }+\lambda ^{\left( 1\right)
}\right) }+e^{i\left( 2\lambda ^{\left( 2\right) }+\lambda ^{\left( 3\right)
}\right) }+e^{i\left( 2\lambda ^{\left( 3\right) }+\lambda ^{\left( 1\right)
}\right) }+e^{i\left( 2\lambda ^{\left( 3\right) }+\lambda ^{\left( 2\right)
}\right) }\right]  \nonumber \\
&&+e^{3i\lambda ^{\left( 1\right) }}+e^{3i\lambda ^{\left( 2\right)
}}+e^{3i\lambda ^{\left( 3\right) }},  \label{eq60}
\end{eqnarray}%
\begin{eqnarray}
D_{001} &=&\left( \tilde{\alpha}_{1}-\tilde{\beta}_{1}\right) \left( \tilde{%
\alpha}_{2}+\omega ^{2}\tilde{\beta}_{2}\right) \left( \tilde{\alpha}%
_{3}+\omega \tilde{\beta}_{3}\right) =2\left( -1+\omega +\omega ^{2}\right)
e^{i\left( \lambda ^{\left( 1\right) }+\lambda ^{\left( 2\right) }+\lambda
^{\left( 3\right) }\right) }  \nonumber \\
&&-\left( \omega +\omega ^{2}-\omega ^{3}\right) \left[ e^{i\left( 2\lambda
^{\left( 1\right) }+\lambda ^{\left( 2\right) }\right) }+e^{i\left( 2\lambda
^{\left( 1\right) }+\lambda ^{\left( 3\right) }\right) }\right] -\left(
1-\omega -\omega ^{2}\right)  \nonumber \\
&&\times \left[ e^{i\left( 2\lambda ^{\left( 2\right) }+\lambda ^{\left(
1\right) }\right) }+e^{i\left( 2\lambda ^{\left( 3\right) }+\lambda ^{\left(
1\right) }\right) }\right] +e^{i\left( 2\lambda ^{\left( 3\right) }+\lambda
^{\left( 2\right) }\right) }+e^{i\left( 2\lambda ^{\left( 2\right) }+\lambda
^{\left( 3\right) }\right) }  \nonumber \\
&&-\omega ^{3}e^{3i\lambda ^{\left( 1\right) }}+e^{3i\lambda ^{\left(
2\right) }}+e^{3i\lambda ^{\left( 3\right) }},  \label{eq61}
\end{eqnarray}%
\begin{eqnarray}
D_{010} &=&\left( \tilde{\alpha}_{1}+\tilde{\beta}_{1}\right) \left( \tilde{%
\alpha}_{2}-\tilde{\beta}_{2}\right) \left( \tilde{\alpha}_{3}+\omega ^{2}%
\tilde{\beta}_{3}\right) =2\omega ^{2}e^{i\left( \lambda ^{\left( 1\right)
}+\lambda ^{\left( 2\right) }+\lambda ^{\left( 3\right) }\right) }-\left[
e^{i\left( 2\lambda ^{\left( 1\right) }+\lambda ^{\left( 2\right) }\right)
}\right.  \nonumber \\
&&\left. +e^{i\left( 2\lambda ^{\left( 1\right) }+\lambda ^{\left( 3\right)
}\right) }\right] +\omega ^{2}\left[ e^{i\left( 2\lambda ^{\left( 2\right)
}+\lambda ^{\left( 1\right) }\right) }+e^{i\left( 2\lambda ^{\left( 3\right)
}+\lambda ^{\left( 1\right) }\right) }\right] +e^{i\left( 2\lambda ^{\left(
3\right) }+\lambda ^{\left( 2\right) }\right) }  \nonumber \\
&&+e^{i\left( 2\lambda ^{\left( 2\right) }+\lambda ^{\left( 3\right)
}\right) }-\omega ^{2}e^{3i\lambda ^{\left( 1\right) }}+e^{3i\lambda
^{\left( 2\right) }}+e^{3i\lambda ^{\left( 3\right) }},  \label{eq62}
\end{eqnarray}%
\begin{eqnarray}
D_{011} &=&\left( \tilde{\alpha}_{1}-\tilde{\beta}_{1}\right) \left( \tilde{%
\alpha}_{2}-\omega ^{2}\tilde{\beta}_{2}\right) \left( \tilde{\alpha}%
_{3}+\omega ^{3}\tilde{\beta}_{3}\right) =2\left( -1-\omega ^{2}+\omega
^{3}\right) e^{i\left( \lambda ^{\left( 1\right) }+\lambda ^{\left( 2\right)
}+\lambda ^{\left( 3\right) }\right) }  \nonumber \\
&&+\left( \omega ^{2}-\omega ^{3}-\omega ^{5}\right) \left[ e^{i\left(
2\lambda ^{\left( 1\right) }+\lambda ^{\left( 2\right) }\right) }+e^{i\left(
2\lambda ^{\left( 1\right) }+\lambda ^{\left( 3\right) }\right) }\right]
-\left( 1+\omega ^{2}-\omega ^{3}\right)  \nonumber \\
&&\times \left[ e^{i\left( 2\lambda ^{\left( 2\right) }+\lambda ^{\left(
1\right) }\right) }+e^{i\left( 2\lambda ^{\left( 3\right) }+\lambda ^{\left(
1\right) }\right) }\right] +e^{i\left( 2\lambda ^{\left( 3\right) }+\lambda
^{\left( 2\right) }\right) }+e^{i\left( 2\lambda ^{\left( 2\right) }+\lambda
^{\left( 3\right) }\right) }  \nonumber \\
&&+\omega ^{5}e^{3i\lambda ^{\left( 1\right) }}+e^{3i\lambda ^{\left(
2\right) }}+e^{3i\lambda ^{\left( 3\right) }},  \label{eq63}
\end{eqnarray}%
\begin{eqnarray}
D_{100} &=&\left( \tilde{\alpha}_{1}+\tilde{\beta}_{1}\right) \left( \tilde{%
\alpha}_{2}+\tilde{\beta}_{2}\right) \left( \tilde{\alpha}_{3}-\tilde{\beta}%
_{3}\right) =2e^{i\left( \lambda ^{\left( 1\right) }+\lambda ^{\left(
2\right) }+\lambda ^{\left( 3\right) }\right) }-2\left[ e^{i\left( 2\lambda
^{\left( 1\right) }+\lambda ^{\left( 2\right) }\right) }\right.  \nonumber \\
&&\left. +e^{i\left( 2\lambda ^{\left( 1\right) }+\lambda ^{\left( 3\right)
}\right) }\right] +2\left[ e^{i\left( 2\lambda ^{\left( 2\right) }+\lambda
^{\left( 1\right) }\right) }+e^{i\left( 2\lambda ^{\left( 3\right) }+\lambda
^{\left( 1\right) }\right) }\right] +e^{i\left( 2\lambda ^{\left( 3\right)
}+\lambda ^{\left( 2\right) }\right) }  \nonumber \\
&&+e^{i\left( 2\lambda ^{\left( 2\right) }+\lambda ^{\left( 3\right)
}\right) }-e^{3i\lambda ^{\left( 1\right) }}+e^{3i\lambda ^{\left( 2\right)
}}+e^{3i\lambda ^{\left( 3\right) }},  \label{eq64}
\end{eqnarray}%
\begin{eqnarray}
D_{101} &=&\left( \tilde{\alpha}_{1}-\tilde{\beta}_{1}\right) \left( \tilde{%
\alpha}_{2}+\omega ^{2}\tilde{\beta}_{2}\right) \left( \tilde{\alpha}%
_{3}-\omega \tilde{\beta}_{3}\right) =-2\left( 1+\omega -\omega ^{2}\right)
e^{i\left( \lambda ^{\left( 1\right) }+\lambda ^{\left( 2\right) }+\lambda
^{\left( 3\right) }\right) }  \nonumber \\
&&+\left( \omega -\omega ^{2}-\omega ^{3}\right) \left[ e^{i\left( 2\lambda
^{\left( 1\right) }+\lambda ^{\left( 2\right) }\right) }+e^{i\left( 2\lambda
^{\left( 1\right) }+\lambda ^{\left( 3\right) }\right) }\right] -\left(
1+\omega -\omega ^{2}\right)  \nonumber \\
&&\times \left[ e^{i\left( 2\lambda ^{\left( 2\right) }+\lambda ^{\left(
1\right) }\right) }+e^{i\left( 2\lambda ^{\left( 3\right) }+\lambda ^{\left(
1\right) }\right) }\right] +e^{i\left( 2\lambda ^{\left( 3\right) }+\lambda
^{\left( 2\right) }\right) }+e^{i\left( 2\lambda ^{\left( 2\right) }+\lambda
^{\left( 3\right) }\right) }  \nonumber \\
&&+\omega ^{3}e^{3i\lambda ^{\left( 1\right) }}+e^{3i\lambda ^{\left(
2\right) }}+e^{3i\lambda ^{\left( 3\right) }},  \label{eq65}
\end{eqnarray}%
\begin{eqnarray}
D_{110} &=&\left( \tilde{\alpha}_{1}+\tilde{\beta}_{1}\right) \left( \tilde{%
\alpha}_{2}-\tilde{\beta}_{2}\right) \left( \tilde{\alpha}_{3}-\omega ^{2}%
\tilde{\beta}_{3}\right) =-2\omega ^{2}e^{i\left( \lambda ^{\left( 1\right)
}+\lambda ^{\left( 2\right) }+\lambda ^{\left( 3\right) }\right) }-\left[
e^{i\left( 2\lambda ^{\left( 1\right) }+\lambda ^{\left( 2\right) }\right)
}\right.  \nonumber \\
&&\left. +e^{i\left( 2\lambda ^{\left( 1\right) }+\lambda ^{\left( 3\right)
}\right) }\right] -\omega ^{2}\left[ e^{i\left( 2\lambda ^{\left( 2\right)
}+\lambda ^{\left( 1\right) }\right) }+e^{i\left( 2\lambda ^{\left( 3\right)
}+\lambda ^{\left( 1\right) }\right) }\right] +e^{i\left( 2\lambda ^{\left(
3\right) }+\lambda ^{\left( 2\right) }\right) }  \nonumber \\
&&+e^{i\left( 2\lambda ^{\left( 2\right) }+\lambda ^{\left( 3\right)
}\right) }+\omega ^{2}e^{3i\lambda ^{\left( 1\right) }}+e^{3i\lambda
^{\left( 2\right) }}+e^{3i\lambda ^{\left( 3\right) }},  \label{eq66}
\end{eqnarray}%
\begin{eqnarray}
D_{111} &=&\left( \tilde{\alpha}_{1}-\tilde{\beta}_{1}\right) \left( \tilde{%
\alpha}_{2}-\omega ^{2}\tilde{\beta}_{2}\right) \left( \tilde{\alpha}%
_{3}-\omega ^{3}\tilde{\beta}_{3}\right) =-2\left( 1+\omega ^{2}+\omega
^{3}\right) e^{i\left( \lambda ^{\left( 1\right) }+\lambda ^{\left( 2\right)
}+\lambda ^{\left( 3\right) }\right) }  \nonumber \\
&&+\left( \omega ^{2}+\omega ^{3}+\omega ^{5}\right) \left[ e^{i\left(
2\lambda ^{\left( 1\right) }+\lambda ^{\left( 2\right) }\right) }+e^{i\left(
2\lambda ^{\left( 1\right) }+\lambda ^{\left( 3\right) }\right) }\right]
-\left( 1+\omega ^{2}+\omega ^{3}\right)  \nonumber \\
&&\times \left[ e^{i\left( 2\lambda ^{\left( 2\right) }+\lambda ^{\left(
1\right) }\right) }+e^{i\left( 2\lambda ^{\left( 3\right) }+\lambda ^{\left(
1\right) }\right) }\right] +e^{i\left( 2\lambda ^{\left( 3\right) }+\lambda
^{\left( 2\right) }\right) }+e^{i\left( 2\lambda ^{\left( 2\right) }+\lambda
^{\left( 3\right) }\right) }  \nonumber \\
&&-\omega ^{5}e^{3i\lambda ^{\left( 1\right) }}+e^{3i\lambda ^{\left(
2\right) }}+e^{3i\lambda ^{\left( 3\right) }}.  \label{eq67}
\end{eqnarray}%
Utilizing ensemble-averaged reduced state, we obtain: 
\begin{eqnarray}
\left\vert \tilde{\Psi}\right\rangle _{F}
&=&\sum\limits_{k=1}^{M}e^{-i\lambda ^{\left( S\right) }}\left\vert \Psi
\right\rangle _{F}=6\left\vert 000\right\rangle -2\left( 1-\omega -\omega
^{2}\right) \left\vert 001\right\rangle +2\omega ^{2}\left\vert
010\right\rangle -2\left( 1+\omega ^{2}-\omega ^{3}\right) \left\vert
011\right\rangle  \nonumber \\
&&+2\left\vert 100\right\rangle -2\left( 1+\omega -\omega ^{2}\right)
\left\vert 101\right\rangle +2\omega ^{2}\left\vert 110\right\rangle
-2\left( 1+\omega ^{2}+\omega ^{3}\right) \left\vert 111\right\rangle .
\label{eq68}
\end{eqnarray}%
In conclusion, $\left\vert \tilde{\Psi}\right\rangle _{F}$ is also the
Fourier transform of $\left\vert \tilde{\Psi}\right\rangle $ for W states.

\section{Conclusion \label{sec5}}

Utilizing pseudorandom phase ensemble model, we propose a transform similar
to quantum Fourier transform. The computational resources required for this
transform is in $O(n^{2})$ similar to quantum Fourier transform, which means
an exponential speedup compared with classical Fourier transform. In the
future, we will utilize this method to implement Shor's algorithm and other
algorithms which need quantum Fourier transform, and then conduct
computer-based simulation of this algorithms to verify the feasibility of
them.

\end{document}